\documentclass[11pt]{article}
\usepackage[margin=2cm]{geometry}
\usepackage{comment}
\usepackage{amsmath,amssymb,extarrows,mathtools,graphicx,subfigure,setspace}
\usepackage{cite}
\usepackage{slashed}
\usepackage{color}
\usepackage{amsmath,amsthm,amsfonts}

\usepackage{fullpage}
\makeatother

\numberwithin{equation}{section}

\newcommand{\be}{\begin{equation}}
\newcommand{\bea}{\begin{eqnarray}}
\newcommand{\eea}{\end{eqnarray}}
\newcommand{\ba}{\begin{array}}
\newcommand{\ea}{\end{array}}
\newcommand{\ee}{\end{equation}}

\def\ba{\begin{array}}
\def\ea{\end{array}}
\def\be{\begin{equation}}
\def\ee{\end{equation}}

\def\-1{^{-1}}

\input{xy}
\xyoption{all}
\xyoption{matrix}

\theoremstyle{plain}

\date{}

\begin{document}

\title {\LARGE {\bf Jacobi-Lie symmetry in WZW model on the \\ Heisenberg Lie group $H_{4}$}}
\vspace{3mm}

\author {  A. Rezaei-Aghdam \hspace{-2mm}{ \footnote{ rezaei-a@azaruniv.edu }} \hspace{2mm}{\small and}\hspace{2mm}
 M. Sephid{ \footnote{s.sephid@azaruniv.edu }}\hspace{2mm}\\
{\small{\em Department of Physics, Faculty of Basic Sciences,
}}\\
{\small{\em  Azarbaijan Shahid Madani University  , 53714-161, Tabriz, Iran  }}}

\maketitle

\vspace{3cm}

\begin{abstract}
We show that the Wess-Zumino-Novikov-Witten (WZW) model on the Heisenberg Lie group $H_{4}$ has Jacobi-Lie symmetry with four dual Lie groups. We construct Jacobi-Lie T-dual sigma models with one of their Jacobi-Lie bialgebra and show that the original model is equivalent to the $H_{4}$ WZW model. The conformality of the dual sigma model up to one-loop order is also investigated.
\end{abstract}
\newpage
\section{\large Introduction}
Target space duality, as a theory, connects different backgrounds in string theory, which makes it an important duality. In T-duality, string backgrounds and their dual fields can be considered as different descriptions of the same physical phenomenon. This duality, at the level of the two-dimensional non-linear sigma models, was started by background fields with abelian and non-abelian isometries (see for a review \cite{GPR1,AAL1}). For abelian case, the dual model can be obtained from the original sigma model and T-duality preserve the symmetries of the original model. But, in non-abelian isometries, this preservation is not possible and target space duality is not reversible \cite{GR1}. A major step to find T-dual sigma models was made by Klim\v{c}ik and \v{S}evera\cite{KS1}. They found duality transformations connecting two non-linear sigma models based on Poisson-Lie groups \cite{Drinfeld1} and called it Poisson-Lie T-duality. This T-duality is based on the interchange between the equations of motion of the original (dual) model and the Bianchi identities of the dual (original) one. In other words, Poisson-Lie T-duality connects two non-linear sigma models when the two corresponding Lie algebras form a Lie bialgebra \cite{Drinfeld1}. But we know that, the Poisson manifolds are a special class of Jacobi manifolds \cite{LichJ}. In fact, we have the very interesting example of Jacobi manifolds such as contact and locally conformal manifolds, which are not Poisson. Therefore, there might be T-dual sigma models where Poisson-Lie T-duality cannot find their corresponding duality transformations. In previous work, we formulated T-duality on Jacobi manifolds and also constructed Jacobi-Lie T-dual sigma models on Lie groups \cite{RS6}.\\
On the other hand, we know that the WZW models \cite{W} play an important role in CFT, since all known conformal field theories can be obtained by decomposition of their Hilbert space. In addition, models where the full symmetry of their action is realized in terms of current algebras are the WZW models with group $G$. But, more attention is given to the construction of WZW models based on non-semi-simple groups\cite{Witten}-\cite{Sfetsos}. These models describe string backgrounds with a target space metric having a covariantly constant null Killing vector\cite{KlTs}. According to \cite{Witten}, we need a non-degenerate invariant bilinear form on the algebra {\bf g} (related to group $G$) which allows constructing a WZW model on non-semi-simple groups. An important four-dimensional Lie group, which benefits from this quadratic form, is the Heisenberg Lie group $H_{4}$.  The WZW model on Heisenberg Lie group, for the first time, was introduced by Kehagias and Meessen \cite{Kehagias}. Analysis of Poisson-Lie symmetry, conformality and D-branes on this model in a different background has been given in \cite{ER9}.\\
In this note, we investigate the Jacobi-Lie T-duality for the WZW model based on the Heisenberg Lie group $H_{4}$. We show that $H_{4}$ WZW model has the Jacobi-Lie symmetry with four dual Lie groups, where Lie group $A_{2}\oplus 2A_{1}$ (found by Poisson-Lie symmetry \cite{ER9}) is one of them.\\
This paper is organized as follows: In section 2, for explanation of the notations and self-containing of the paper, we review some aspects of the Jacobi-Lie symmetry on group manifolds \cite{RS6}. In section 3, we obtain the WZW model on the Heisenberg Lie group $H_{4}$ similar to Ref.\cite{ER9}. Then, we apply Jacobi-Lie symmetry condition on this action and show that $H_{4}$ WZW  model has this symmetry with four dual Lie groups, where obtained dual Lie group $A_{2}\oplus 2A_{1}$ \cite{ER9} is one of them. In section 4, we first briefly review the construction of Jacobi-Lie T-dual sigma models on Lie groups. Then, due to the fact that the classical $r$-matrices on original and dual groups must be existent in our formalism, we construct Jacobi-Lie T-dual sigma models only on Jacobi-Lie bialgebra $((h_{4},\phi_{0}),({\cal A}^{-1,1}_{4,5}.i,X_{0}))$ and we have shown that the original sigma model is equivalent to the $H_{4}$WZW model. Analysis of conformal invariance for Jacobi-Lie T-dual sigma models is given in section 5. In this section, we first write sigma model action with dilaton field $\Phi$ in the presence of $\sigma$-function. Then, we give the one-loop $\beta$-function equations for this model. In this way, we have shown the conformal invariance of the dual sigma model for a pair $((h_{4},\phi_{0}),({\cal A}^{-1,1}_{4,5}.i,X_{0}))$ and we have obtained the general form of the dilaton fields. Some discussions and concluding remarks are given at the end.
\section{\large Review of the Jacobi-Lie symmetry}
In order to introduce notational conventions, let us start this section with a short review of the Jacobi-Lie symmetry \cite{RS6}. Consider a non-linear sigma model on a manifold $M$ with metric $g_{\mu\nu}$ and antisymmetric tensor field $b_{\mu\nu}$ in the presence of a $C^{\infty}$-function $\sigma$ as follows
\begin{equation}\label{1}
S=\frac{1}{2}\int_{\Sigma}{d\xi^{+}\wedge d\xi^{-}~e^{\sigma(x)}{\large (}{g}_{\mu\nu}(x)+{b}_{\mu\nu}(x){\large )}~ \partial_{+}x^{\mu}\partial_{-}x^{\nu}}=\frac{1}{2}\int_{\Sigma}{d\xi^{+}\wedge d\xi^{-}~e^{\sigma(x)}{\cal E}_{\mu\nu}(x)~ \partial_{+}x^{\mu}\partial_{-}x^{\nu}},
\end{equation}
where $\partial_{\pm}$ are the derivatives with respect to standard light-cone variables $\xi^{\pm}$ on the worldsheet $\Sigma$ and $x^{\mu}$ ($\mu=1,..,dimM$) are coordinates on $M$. Suppose that Lie group $G$ (with Lie algebra ${\bf g}$\footnote{with bases $\{X_{i}\}$ and structure constants $f_{ij}\hspace{0cm}^{k}$}) acts freely on $M$ from the right by the left-invariant vector fields $v_{i}\hspace{0cm}^{\mu}$. Then, the Hodge star of Noether's current one-forms corresponding to this action has the following form \cite{RS6}
\begin{equation}\label{2}
\star J_{i}=e^{\sigma({x})}({\cal E}_{\mu\gamma}~v_{i}\hspace{0cm}^{\gamma}~ \partial_{+}x^{\mu}d\xi^{+}-{\cal E}_{\gamma\nu}~v_{i}\hspace{0cm}^{\gamma}~ \partial_{-}x^{\nu}d\xi^{-}).
\end{equation}
Now, if the 1-forms $\star J_{i}$ on the extremal surface $x^{\mu}(\xi^{+},\xi^{-})$ are not closed and they satisfy the following generalized Maurer-Cartan equation\cite{RS6}
\begin{equation}\label{3}
d \star J_{i}=-\frac{1}{2}e^{-\sigma}({\tilde{f}^{jk}}\hspace{0cm}_{i}-\alpha^{j}\delta^{k}
\hspace{0cm}_{i} +\alpha^{k}\delta^{j}\hspace{0cm}_{i}) \star J_{j} \wedge \star J_{k},
\end{equation}
and using the equations of motion related to \eqref{1}, the Jacobi-Lie symmetry condition on the background matrix ${\cal E}_{\mu\nu}=g_{\mu\nu}+b_{\mu\nu}$ can be formulated as\cite{RS6}
\begin{equation}\label{4}
({\cal L}_{\phi_{0}})_{v_{i}}{\cal E}_{\mu\nu}=({\tilde{f}^{jk}}\hspace{0cm}_{i}-\alpha^{j}\delta^{k}
\hspace{0cm}_{i} +\alpha^{k}\delta^{j}\hspace{0cm}_{i}) v_{j}\hspace{0cm}^{\lambda} v_{k}\hspace{0cm}^{\eta} {\cal E}_{\mu\eta} {\cal E}_{\lambda\nu},
\end{equation}
where $\tilde{f}^{jk}\hspace{0cm}_{i}$ are structure constants of the dual Lie algebra ${ \bf\tilde g}$ and $\alpha^{i}$ are coefficients of $X_{0}=d{\tilde \sigma}\in {\bf g}$ (${\tilde \sigma}$ is a function on the dual manifold $\tilde M$). In this relation, $\phi_{0}=d\sigma\in\Omega^{1}(M)$ is a 1-cocycle on ${\bf g}$ with values in ${\bf \tilde g}$ \footnote{$\phi_{0}=\beta_{i}{\tilde X^{i}}$ where $\beta_{i}\equiv v_{i}\hspace{0cm}^{\lambda}\partial_{\lambda}\sigma$ and $\{{\tilde X^{i}}\}$ are the basis of ${\bf \tilde g}$ \cite{RS6}} and $\phi_{0}$-Lie derivative has the following definition \cite{IM1}
\begin{equation}\label{5}
({\cal L}_{\phi_{0}})_{v_{i}}{\cal E}_{\mu\nu}={\cal L}_{v_{i}}{\cal E}_{\mu\nu}+<\phi_{0},v_{i}>{\cal E}_{\mu\nu}.
\end{equation}
The Jacobi-Lie symmetry \eqref{4} is a generalization of the Poisson-Lie symmetry (with $\sigma=0$ and $\alpha^{i}=0$)\cite{KS1}. The integrability condition for Jacobi-Lie symmetry \eqref{4}, using 1-cocycle condition for $\phi_{0}$, i.e.,
\begin{equation}\label{6}
\beta_{k}{f}_{ij}\hspace{0cm}^{k}=0,
\end{equation}
gives compatibility between structure constants of Lie algebras ${\bf g}$ and ${\bf \tilde g}$ and also coefficient constants $\alpha^{i}$ and $\beta_{j}$ as follows
$$
\hspace{-.3cm}{f}_{ij}\hspace{0cm}^k{\tilde{f}^{mn}}\hspace{0cm}_{k}-{f}_{ik}\hspace{0cm}^m{\tilde{f}^{kn}}\hspace{0cm}_{j} -\\
{f}_{ik}\hspace{0cm}^n{\tilde{f}^{mk}}\hspace{0cm}_{j}-{f}_{kj}\hspace{0cm}^m{\tilde{f}^{kn}}\hspace{0cm}_{i}-\\
{f}_{kj}\hspace{0cm}^n{\tilde{f}^{mk}}\hspace{0cm}_{i}+\beta_{i}{\tilde{f}^{mn}}\hspace{0cm}_{j}-\beta_{j}{\tilde{f}^{mn}}\hspace{0cm}_{i}+\alpha^{m}{f}_{ij}\hspace{0cm}^n-\alpha^{n}{f}_{ij}\hspace{0cm}^m\\
$$
\begin{equation}\label{7}
+(\alpha^{k}{f}_{ik}\hspace{0cm}^m-\alpha^{m}\beta_{i})\delta_{j}\hspace{0cm}^{n}
-(\alpha^{k}{f}_{jk}\hspace{0cm}^m-\alpha^{m}\beta_{j})\delta_{i}\hspace{0cm}^{n}-(\alpha^{k}{f}_{ik}\hspace{0cm}^n-\alpha^{n}\beta_{i})\delta_{j}\hspace{0cm}^{m}
+(\alpha^{k}{f}_{jk}\hspace{0cm}^n-\alpha^{n}\beta_{j})\delta_{i}\hspace{0cm}^{m}=0.
\end{equation}
The relation \eqref{7} is the first condition of the Jacobi-Lie bialgebra $(({\bf{g}},\phi_{0}),({\tilde{\bf g}},X_{0}))$ \cite{RS4}.\\
In the same way, one can consider the following action for the dual model with the background matrix ${\cal \tilde E}_{\mu\nu}=\tilde g_{\mu\nu}+\tilde b_{\mu\nu}$ and $\tilde\sigma$-function on target space $\tilde{M}$
\begin{equation}\label{7'}
{\tilde S}=\frac{1}{2}\int{d\xi^{+}\wedge d\xi^{-}~e^{\tilde \sigma(\tilde x)}{\large (}{\tilde g}_{\mu\nu}(\tilde x)+{\tilde b}_{\mu\nu}(\tilde x){\large )}~ \partial_{+}{\tilde x}^{\mu}\partial_{-}{\tilde x}^{\nu}},
\end{equation}
and with regard to the following generalized Maurer-Cartan equation
\begin{equation}\label{8}
d \star \tilde J^{i}=-\frac{1}{2}e^{-\tilde \sigma}({{f}_{jk}}\hspace{0cm}^{i}-\beta_{j}\delta_{k}
\hspace{0cm}^{i} +\beta_{k}\delta_{j}\hspace{0cm}^{i}) \star \tilde J^{j} \wedge \star \tilde J^{k},
\end{equation}
for the Neother's currents $\tilde J^{i}$ (related to the free action of Lie group $\tilde G$ on manifold $\tilde M$); we have the Jacobi-Lie symmetry for the dual model \eqref{7'} as
\begin{equation}\label{9}
({\cal {\tilde L}}_{X_{0}})_{\tilde v^{i}}{\cal \tilde E}_{\mu\nu}=({{f}_{jk}}\hspace{0cm}^{i}-\beta_{j}\delta_{k}
\hspace{0cm}^{i} +\beta_{k}\delta_{j}\hspace{0cm}^{i}) \tilde v^{j}\hspace{0cm}^{\lambda} \tilde v^{k}\hspace{0cm}^{\eta} {\cal \tilde E}_{\mu\eta} {\cal \tilde E}_{\lambda\nu}.
\end{equation}
In the same way, the integrability condition for \eqref{9} is equivalent to the following relations \cite{RS6}
\begin{equation}\label{10}
\alpha^{k}{\tilde{f}^{mn}}\hspace{0cm}_{k}=0,
\end{equation}
and
$$
\hspace{-.3cm}{f}_{ij}\hspace{0cm}^k{\tilde{f}^{mn}}\hspace{0cm}_{k}-{f}_{ik}\hspace{0cm}^m{\tilde{f}^{kn}}\hspace{0cm}_{j} -\\
{f}_{ik}\hspace{0cm}^n{\tilde{f}^{mk}}\hspace{0cm}_{j}-{f}_{kj}\hspace{0cm}^m{\tilde{f}^{kn}}\hspace{0cm}_{i}-\\
{f}_{kj}\hspace{0cm}^n{\tilde{f}^{mk}}\hspace{0cm}_{i}+\beta_{i}{\tilde{f}^{mn}}\hspace{0cm}_{j}-\beta_{j}{\tilde{f}^{mn}}\hspace{0cm}_{i}+\alpha^{m}{f}_{ij}\hspace{0cm}^n-\alpha^{n}{f}_{ij}\hspace{0cm}^m\\
$$
\begin{equation}\label{11}
+(\beta_{k}{\tilde f}^{mk}\hspace{0cm}_i-\alpha^{m}\beta_{i})\delta_{j}\hspace{0cm}^{n}
-(\beta_{k}{\tilde f}^{mk}\hspace{0cm}_j-\alpha^{m}\beta_{j})\delta_{i}\hspace{0cm}^{n}-(\beta_{k}{\tilde f}^{nk}\hspace{0cm}_i-\alpha^{n}\beta_{i})\delta_{j}\hspace{0cm}^{m}
+(\beta_{k}{\tilde f}^{nk}\hspace{0cm}_j-\alpha^{n}\beta_{j})\delta_{i}\hspace{0cm}^{m}=0.
\end{equation}
The relations \eqref{10} and \eqref{11} set the 1-cocycle condition on $X_{0}$ and the first condition of the Jacobi-Lie bialgebra $(({\tilde{\bf g}},X_{0}),({\bf{g}},\phi_{0}))$, respectively \cite{RS4}. The subtraction of the relation \eqref{7} and \eqref{11}, gives the symmetrizing condition of the Jacobi-Lie bialgebras with respect to  pairs $({\bf{g}},\phi_{0})$ and $({\tilde{\bf g}},X_{0})$, i.e.,
\begin{equation}\label{12}
\alpha^{k}{f}_{ik}\hspace{0cm}^{m}-\beta_{k}{\tilde{f}^{mk}}\hspace{0cm}_{i}=0.
\end{equation}
Now, if 1-cocycles $X_{0}$ and $\phi_{0}$ are satisfied in the following condition
\begin{equation}\label{13}
\alpha^{i}\beta_{i}=0,
\end{equation}
we obtain the last condition of Jacobi-Lie bialgebras $(({\bf{g}},\phi_{0})({\tilde{\bf g}},X_{0}))$ and $(({\tilde{\bf g}},X_{0}),({\bf{g}},\phi_{0}))$ \cite{RS4,IM2}.
Therefore, if Lie algebras ${\bf g}$ and ${\bf \tilde g}$ and 1-cocycles $X_{0}$ and $\phi_{0}$ are satisfied in relations \eqref{6},\eqref{7},\eqref{10},\eqref{12} and \eqref{13}, we have Jacobi-Lie bialgebra $(({\bf{g}},\phi_{0}),({\tilde{\bf g}},X_{0}))$ \cite{RS4}. Consequently, the invariance of sigma models $S$ and $\tilde S$ on manifolds $M$ and $\tilde M$, under the free action of $G$ on $M$ and ${\tilde G}$ on ${\tilde M}$, respectively, is equivalent to the Jacobi-Lie bialgebra structure \cite{RS6}.
\section{\large Jacobi-Lie symmetry in the  WZW model on the Heisenberg Lie group $H_4$}
Writing the WZW action corresponding to non-semi-simple groups is not straightforward, since their Killing metric $\Omega_{ij}=f_{ik}\hspace{0cm}^{l}f_{jl}\hspace{0cm}^{k}$ is degenerate \cite{Witten}-\cite{KlTs}. Nappi and Witten in \cite{Witten} resolve this problem by considering another symmetric non-degenerate ad-invariant bilinear form $\Omega_{ij}=<X_{i},X_{j}>$ in the following relation
\begin{equation}\label{14}
<X_{i},[X_{j},X_{k}]>=<[X_{i},X_{j}],X_{k}>,
\end{equation}
for a Lie algebra with generators $X_{i}$ and structure constants $[X_{i},X_{j}]=f_{ij}\hspace{0cm}^{k}X_{k}$. The WZW action on Lie group $G$ by this quadratic form is written as \cite{Witten} \begin{eqnarray}\label{15}
S_{_{WZW}}(g) &=&  \frac{K}{4\pi} \int_{\Sigma} d\xi^{+}\wedge d\xi^{-}\;
L^{\hspace{-0.5mm}i}_{+}\;{\Omega}_{ij}\;
L^{\hspace{-0.5mm}j}_{-}
+\frac{K}{24\pi} \int_{B} d^3 \xi~
\varepsilon^{ \gamma \alpha \beta}
L^{\hspace{-0.5mm}i}_{\gamma}
\;{\Omega}_{il}\;L^{\hspace{-0.5mm}j}_{\alpha}
\;f_{jk}^{~~l}~ L^{\hspace{-0.5mm}k}_{\beta},
\end{eqnarray}
where $\Sigma$ is two-dimensional worldsheet and $B$ a three-dimensional manifold with boundary $\partial B=\Sigma$. In this action, $L^{\hspace{-0.5mm}i}_{\alpha}$'s are components of left-invariant 1-forms defined via $g^{-1} \partial_{\alpha} g=L^{\hspace{-0.5mm}i}_{\alpha}X_{i}$ which $g$ is a map of $\Sigma$ to $G$.\\
The aim of this paper, is analyzing the WZW model on the Heisenberg Lie group that have Jacobi-Lie symmetry. Before constructing the model, we need the structure constants of Lie algebra $h_{4}$ of the Lie group $H_{4}$. The oscillator Lie algebra $h_{4}$ consists of four generators $\{a, a^{\dagger}, N=aa^{\dagger}, M\}$ with the following commutation relations
\begin{eqnarray}\label{16}
[N , a^{\dagger}]~=~a^{\dagger},~~~~~[N , a]=-a,~~~~~[a , a^{\dagger}]~=~M.
\end{eqnarray}
As mentioned, to construct a WZW model on a non-semi-simple Lie group $G$, we need ad-invariant symmetric bilinear form $\Omega_{ij}$. One can obtain this quadratic form by the following matrix form using the relation \eqref{14} \footnote{t stands for transpose}
\begin{equation}\label{17}
{\cal X}_{i}\Omega+({\cal X}_{i} \Omega)^{t}=0,
\end{equation}
where $({\cal X}_{i})_{j}\hspace{0cm}^{k}=-f_{ij}\hspace{0cm}^{k}$ is the adjoint representation of Lie algebra ${\bf g}$. Solving Eq.\eqref{17}, one can obtain non-degenerate form $\Omega_{ij}$, as follows \cite{ER9},
\begin{eqnarray}\label{18}
\Omega_{ij}=\left(
              \begin{array}{cccc}
                0 & 0 & 0 & -\kappa \\
                0 & 0 & \kappa & 0 \\
                0 & \kappa & 0 & 0 \\
                -\kappa & 0 & 0 & \kappa' \\
              \end{array}
              \right),\qquad   \kappa\in \Re-\{0\},\;\;\; \kappa'\in \Re.
\end{eqnarray}
To write the explicit form of the action \eqref{15}, we need the parametrization of the Lie group $H_{4}$. There exists various parameterizations for Lie group $H_{4}$ (for example \cite{Kehagias, ER9, RS2}). The convenient parametrization for this paper is the same as Ref.\cite{ER9} with coordinates $x^{\mu}=\{x,y,u,v\}$ as
\begin{eqnarray}\label{19}
g\;=\; e^{v X_4} ~ e^{u X_3} ~ e^{x X_1}~ e^{y X_2},
\end{eqnarray}
where we have used the new generators $\{X_{1},X_{2},X_{3},X_{4}\}$ instead of $\{N,a,a^{\dagger},M\}$. The $L^{\hspace{-0.5mm}i}_{\alpha}$'s are found to be
\begin{eqnarray}\label{20}
L_{\alpha}=\partial_{\alpha} x\;X_{1}+(y \partial_{\alpha} x + \partial_{\alpha} y)\;X_{2}+e^{x}~\partial_{\alpha} u\;X_{3}+(ye^{x}~ \partial_{\alpha} u + \partial_{\alpha} v)\;X_{4}.
\end{eqnarray}
Using \eqref{18} (with $\kappa'=0$) and \eqref{20}, the integrate terms in \eqref{15} are obtained as follows
$$
L^{\hspace{-0.5mm}i}_{+}\;{\Omega}_{ij}\;
L^{\hspace{-0.5mm}j}_{-}=\kappa\Big\{-\partial_+x \partial_-v -\partial_+v \partial_-x
+e^x\Big(\partial_+y \partial_-u+\partial_+u \partial_-y\Big)\Big\},
$$
\vspace{-6mm}
\begin{eqnarray}\label{21}
\varepsilon^{ \gamma \alpha \beta}
L^{\hspace{-0.5mm}i}_{\gamma}
\;{\Omega}_{il}\;L^{\hspace{-0.5mm}j}_{\alpha}
\;f_{jk}^{~~l}~ L^{\hspace{-0.5mm}k}_{\beta}=6\;\kappa\;\varepsilon^{\gamma \alpha \beta} e^{x}\;\partial_{\alpha}x\;\partial_{\beta}y\;\partial_{\gamma}u.
\end{eqnarray}
Therefore, the WZW action on the Lie group $H_{4}$ (by choosing $\varepsilon^{+-}=-\varepsilon^{-+}=1$) is written as
\begin{eqnarray}\label{22}
S_{_{WZW}}=\frac{\kappa K}{4\pi}\int d\xi^{+}\wedge d\xi^{-}\Big\{-\partial_+x\partial_-v-\partial_+v\partial_-x
+e^x\Big(\partial_+y\partial_-u+\partial_+u\partial_-y\nonumber\\
+y\partial_+u\partial_-x-y\partial_+x\partial_-u\Big)\Big\}.
\end{eqnarray}
The above action on Lie group $H_{4}$, for the first time, was constructed in \cite{ER9}. In that work, it was shown that \eqref{22} has the Poisson-Lie symmetry only when the dual Lie group is $A_{2}\oplus2A_{1}$. As follows, we will consider the Jacobi-Lie symmetry on $H_{4}$WZW model and show that dual Lie group $A_{2}\oplus2A_{1}$  is one of the found dual Lie groups by Jacobi-Lie symmetry.\\
By identifying the action \eqref{22} with the sigma model action \eqref{1}, we can read off the background matrix as
\begin{eqnarray}\label{24}
e^{\sigma(x)} {\cal E}_{\mu \nu}~=~\frac{\kappa K}{2\pi}\left( \begin{tabular}{cccc}
              $0$ & $0$ & $-y e^x$ & $-1$\\
              $0$ & $0$ & $e^x$ & $0$ \\
              $y e^x$ & $e^x$ & $0$ & $0$ \\
              $-1$ & $0$ & $0$ & $0$\\
                \end{tabular} \right).
\end{eqnarray}
In order to investigate the Jacobi-Lie symmetry in WZW model \eqref{22}, we need the left-invariant vector fields on Lie group $H_{4}$. Substituting relation \eqref{20} into equation $<V_{i},L^{j}>=\delta_{\hspace{-0.5mm}i}^{j}$ (using $L_{\alpha}=L^{\hspace{-0.5mm}i}_{\alpha} X_{i}$), $V_{i}$'s are obtained to be
\begin{eqnarray}\label{25}
{V}_{1}=\frac{\partial}{\partial x}-y \frac{\partial}{\partial y},~~~~~{V}_{2}= \frac{\partial}{\partial y},~~~~~
{V}_{3}=e^{-x}~\frac{\partial}{\partial u}-y \frac{\partial}{\partial v},~~~~~{V}_{4}=\frac{\partial}{\partial v}.
\end{eqnarray}
Now, we calculate $\phi_{0}$-Lie derivative of relation \eqref{24} with respect to \eqref{25} and then we put these results into \eqref{4}. Then, by substituting obtained $\alpha$,$\beta$ and $\tilde f$ in equations \eqref{10}-\eqref{12} and \eqref{13}, the non-zero commutation relations of the dual pair to the Heisenberg Lie algebra $h_{4}$ and 1-cocycles $X_{0}$ and $\phi_{0}$ are found to be\\

\hspace{-5.5mm}i)~$((h_{4},0),({\cal A}_{2}\oplus2{\cal A}_{1},0))$
\begin{equation}\label{26}
[{\tilde X}^2 , {\tilde X}^4]~=~{\tilde X}^2
\end{equation}
ii)~$((h_{4},0),({\cal V}\oplus \Re.i,X_{4}))$
\begin{equation}\label{27}
[{\tilde X}^1 , {\tilde X}^4]~=-{\tilde X}^1~~,~~[{\tilde X}^3 , {\tilde X}^4]~=-{\tilde X}^3
\end{equation}
iii)~$((h_{4},0),({\cal A}^{a,a}_{4,5}.i,\frac{a}{a-1}X_{4}))$
\begin{equation}\label{28}
[{\tilde X}^1 , {\tilde X}^4]~=-\frac{a}{a-1}{\tilde X}^1~~,~~[{\tilde X}^2 , {\tilde X}^4]~=-\frac{1}{a-1}{\tilde X}^2~~,~~[{\tilde X}^3 , {\tilde X}^4]~=-\frac{a}{a-1}{\tilde X}^3
\end{equation}
iv)~$((h_{4},0),({\cal A}^{a,1}_{4,5}.i,\frac{1}{1-a}X_{4}))$
\begin{equation}\label{29}
[{\tilde X}^1 , {\tilde X}^4]~=\frac{1}{a-1}{\tilde X}^1~~,~~[{\tilde X}^2 , {\tilde X}^4]~=\frac{a}{a-1}{\tilde X}^2~~,~~[{\tilde X}^3 , {\tilde X}^4]~=\frac{1}{a-1}{\tilde X}^3
\end{equation}
As expected, the first class of the Jacobi-Lie bialgebras on $h_{4}$ is the Lie bialgebra $(h_{4},{\cal A}_{2}\oplus2{\cal A}_{1})$\footnote{Lie algebra ${\cal A}_{2}\oplus2{\cal A}_{1}$ (where ${\cal A}_{1}$ is one-dimensional Lie algebra) is the same as Lie algebra $III\oplus \Re$ in the classification of four-dimensional Lie algebras \cite{Patera}} with $\alpha^{i}=\beta_{i}=0$. This dual pair, for the first time, was obtained in \cite{ER9} from the Poisson-Lie symmetry on $H_{4}$WZW. Hence, we denote that the Poisson-Lie symmetry can be obtained from the Jacobi-Lie symmetry when $X_{0}=\phi_{0}=0$ (or $\sigma={\tilde\sigma}=0$). Furthermore, from the Jacobi-Lie symmetry of WZW model on the Heisenberg Lie group $H_{4}$, one can obtain three dual Lie algebras $({\cal V}\oplus \Re.i)$, $({\cal A}^{a,a}_{4,5}.i)$ and $({\cal A}^{a,1}_{4,5}.i)$ with $X_{0}\neq0$.\\ In the next section, we will try to construct the Jacobi-Lie T-dual sigma models which are associated with these Jacobi-Lie bialgebras.
\section{\large Jacobi-Lie T-dual sigma models on the Heisenberg Lie group $H_{4}$ and its duals}
In the previous work\cite{RS6}, we introduced the formalism of the Jacobi-Lie T-dual sigma models with Lie group as a target space. As mentioned in \cite{RS6}, according to the duality between ${\bf g}$ and ${\bf \tilde g}$
\begin{equation}\label{30}
<X_{i},{\tilde X}^{j}>=\delta_{i}\hspace{0cm}^{j}
\end{equation}
and the commutation relations between  $\{X_{i}\}$ and $\{\tilde{X}^{j}\}$ as
\begin{equation}\label{31}
[X_i , \tilde{X}^j] =({\tilde{f}^{jk}}\hspace{0cm}_{i}+\frac{1}{2}\alpha^{k}\delta_{i}\hspace{0cm}^{j}-\alpha^{j}\delta_{i}\hspace{0cm}^{k})X_k +({f}_{ki}\hspace{0cm}^{j}-\frac{1}{2}\beta_{k}\delta_{i}\hspace{0cm}^{j}+\beta_{i}\delta_{k}\hspace{0cm}^{j}) \tilde{X}^k,
\end{equation}
the Jacobi-Lie T-dual sigma models $S$ and $\tilde S$ related to the 2n-dimensional vector space ${\cal D}={\bf g}\oplus{\bf \tilde g}$ ($n=dim{\bf g}=dim{\bf \tilde g}$), have the following definitions
\begin{eqnarray}\label{32}
S=\frac{1}{2}\int{d\xi^{+}} \wedge d\xi^{-}~[E^{n+}(g)]_{ij} ~(g^{-1} \partial_{+} g)^{i}~(g^{-1} \partial_{-} g)^{j},
\end{eqnarray}
\vspace{-6mm}
\begin{eqnarray}\label{33}
\tilde S=\frac{1}{2}\int{d\xi^{+}} \wedge d\xi^{-}~[\tilde E^{n+}(\tilde g)]\hspace{0cm}^{ij} ~(\tilde g^{-1} \partial_{+} \tilde g)_{i}~(\tilde g^{-1} \partial_{-} \tilde g)_{j},
\end{eqnarray}
where background matrices $E^{n+}(g)$ and $\tilde E^{n+}(\tilde g)$ are defined by \cite{RS6}
\begin{eqnarray}\label{34}
\hspace{9mm}E^{n+}(g)=(a(g)+E^{+}({e}) a^{-t}(g) \Lambda(g))^{-1} E^{+}({e}) a^{-t}(g),
\end{eqnarray}
\begin{equation}\label{35}
{\tilde E}^{n+}({\tilde g})=({\tilde a}({\tilde g})+{\tilde E}^{+}({\tilde e}) {\tilde a}^{-t}({\tilde g}) {\tilde \Lambda} ({\tilde g}))^{-1} {\tilde E}^{+}({ \tilde e}) {\tilde a}^{-t}({\tilde g}),
\end{equation}
such that $a(g)$ and ${\tilde a (\tilde g)}$ are given as follows
\begin{equation}\label{36}
\forall g\in{G},~{\tilde g}\in{\tilde G}~~~~~~~~~~~~g^{-1}X_{i}~g= {a(g)}_{i}\hspace{0cm}^{j}~X_{j}~~~,~~~{\tilde g}^{-1}{\tilde X}^{i}~{\tilde g}= {{\tilde a}({\tilde g})}^{i}\hspace{0cm}_{j}~{\tilde X}^{j},
\end{equation}
and bi-vector fields $\Lambda(g)$ and ${\tilde \Lambda(\tilde g)}$ on the Lie groups $G$ and ${\tilde G}$ have the following forms
\begin{eqnarray}\label{37}
\Lambda^{ij}=r^{mn}~a(g)_{m}\hspace{0cm}^{i}~a(g)_{n}\hspace{0cm}^{j} - e^{-\sigma} r^{ij},
\end{eqnarray}
\vspace{-9mm}
\begin{eqnarray}\label{38}
{\tilde\Lambda}_{ij}={\tilde r}_{mn}~{\tilde a}({\tilde g})^{m}\hspace{0cm}_{i}~{\tilde a}({\tilde g})^{n}\hspace{0cm}_{j} - e^{-{\tilde\sigma}} {\tilde r}_{ij},
\end{eqnarray}
where $r$ and ${\tilde r}$ are classical $r$-matrices on Jacobi-Lie bialgebras $(({\bf{g}},\phi_{0}),({\tilde{\bf g}},X_{0}))$ and $(({\tilde{\bf g}},X_{0}),({\bf{g}},\phi_{0}))$ \cite{RS5}. Furthermore, $E^{+}({e})$ and ${\tilde E}^{+}({\tilde e})$ are constant matrices at the unit element of $G$ and ${\tilde G}$, respectively, and are related to each other, as
\begin{eqnarray}\label{39}
E^{+}({e}){\tilde E^{+}}({\tilde e})={\tilde E^{+}}({\tilde e})E^{+}({e})=I,
\end{eqnarray}
which is obtained considering non-degeneracy of the maps $E(g):{\bf g}\rightarrow{\bf \tilde g}$ and ${\tilde E}({\tilde g}):{\bf \tilde g}\rightarrow {\bf g}$.\\
 According to \eqref{37} and \eqref{38}, the classical $r$ and ${\tilde r}$-matrices on Jacobi-Lie bialgebras  are the essential tools in our method to construct Jacobi-Lie T-dual sigma models. This restriction comes from the absence of ad-invariance on the inner product \eqref{30} in Jacobi-Lie bialgebras. According to our finding in the previous section about Jacobi-Lie bialgebras for $H_{4}$ related to the Jacobi-Lie symmetry of WZW model, only $((h_{4},0),({\cal A}^{a,a}_{4,5}.i,\frac{a}{a-1}X_{4}))$ and $((h_{4},0),({\cal A}^{a,1}_{4,5}.i,\frac{1}{1-a}X_{4}))$ with $a=-1$ are bi-$r$-matrix\footnote{Coboundary Jacobi-Lie bialgebras with classical $r$ and ${\tilde r}$ matrices.} and these Jacobi-Lie bialgebras (for $a=-1$) are equivalent\cite{RS4}. In this way, we will construct the Jacobi-Lie T-dual sigma models which are associated with the Jacobi-Lie bialgebras $((h_{4},0),({\cal A}^{-1,1}_{4,5}.i,\frac{1}{2}X_{4}))$. We will show that the original sigma model on this Jacobi-Lie bialgebra is equivalent to $H_{4}$WZW model.
\subsection{\large Jacobi-Lie T-dual sigma models on the Jacobi-Lie bialgebra $((h_4,\phi_{0})~({\cal A}^{-1,1}_{4,5}.i,X_{0}))$}
Consider Lie algebra ${\bf g}=h_{4}$ defined by commutation relations
\begin{eqnarray}\label{40}
[X_{1},X_{2}]=X_{2},~~~~~[X_{1},X_{3}]=-X_{3},~~~~~[X_{2},X_{3}]=-X_{4}.
\end{eqnarray}
and its dual ${\bf \tilde g}={\cal A}^{-1,1}_{4,5}.i$ which is generated by relation \eqref{29} (with $a=-1$), in the presence of 1-cocycles $X_{0}=\frac{1}{2}X_{4}$ and $\phi_{0}=0$ ($\beta_{1}=\beta_{2}=\beta_{3}=\beta_{4}=0$ and $\alpha^{1}=\alpha^{2}=\alpha^{3}=0, \alpha^{4}=\frac{1}{2}$) and functions $\sigma=0$ and ${\tilde\sigma}=\frac{\tilde v}{2}$. Using \eqref{31}, we have the following non-zero commutation relations on vector space ${\cal D}={\bf g}\oplus{\bf \tilde g}$ with eight generators $\{X_{1},..,X_{4},{\tilde X}^{1},..,{\tilde X}^{4}\}$\footnote{Note that for this Jacobi-Lie bialgebra the vector space ${\cal D}$ is not a Lie algebra, i.e. the commutation relations \eqref{41} do not satisfy the Jacobi identity.}

\begin{eqnarray*}
\hspace{-1.4cm}[X_{1},{\tilde X}^{1}]=-\frac{X_{4}}{4}~~~~~~~~~~~[X_{1},{\tilde X}^{2}]=-{\tilde X}^{2}~~~~~~~~~~~~[X_{1},{\tilde X}^{3}]={\tilde X}^{3}
\end{eqnarray*}
\vspace{-1cm}
\begin{eqnarray*}
[X_{2},{\tilde X}^{2}]=\frac{3}{4}X_{4}+{\tilde X}^{1}~~~~[X_{2},{\tilde X}^{4}]=-X_{2}+{\tilde X}^{3}~~~~[X_{3},{\tilde X}^{3}]=-\frac{X_{4}}{4}-{\tilde X}^{1}
\end{eqnarray*}
\vspace{-1cm}
\begin{eqnarray}\label{41}
\hspace{-5cm}[X_{3},{\tilde X}^{4}]=-{\tilde X}^{2}~~~~~~~~~~~[X_{4},{\tilde X}^{4}]=-\frac{X_{4}}{4}.
\end{eqnarray}
\subsubsection{\normalsize The original model}
In order to construct sigma model \eqref{32}, at first, by a direct application of the classical $r$-matrix formula for coboundary Jacobi-Lie bialgebra $((h_{4},0),({\cal A}^{-1,1}_{4,5}.i,\frac{1}{2}X_{4}))$ we have\cite{RS5}
\begin{eqnarray}\label{42}
r^{ij}=\left(
              \begin{array}{cccc}
                0 & 0 & 0 & \frac{1}{2} \\
                0 & 0 & \frac{1}{2} & 0 \\
                0 & -\frac{1}{2} & 0 & 0 \\
                -\frac{1}{2} & 0 & 0 & 0 \\
              \end{array} \right).
\end{eqnarray}
Also, by using the same parametrization \eqref{19} in \eqref{36}, we find that
\begin{eqnarray}\label{43}
a(g)_{i}\hspace{0cm}^{j}=\left(
              \begin{array}{cccc}
                1 & y & -ue^{x} & -yue^{x} \\
                0 & e^{-x} & 0 & -u \\
                0 & 0 & e^{x} & ye^{x} \\
               0 & 0 & 0 & 1 \\
              \end{array} \right).
\end{eqnarray}
Inserting \eqref{42} and \eqref{43} and $\sigma=0$ in \eqref{37}, one can obtain bi-vector $\Lambda(g)$, as
\begin{eqnarray}\label{44}
\Lambda(g)^{ij}=\left(
              \begin{array}{cccc}
                0 & 0 & 0 & 0 \\
                0 & 0 & 0 & y \\
                0 & 0 & 0 & 0 \\
               0 & -y & 0 & 0 \\
              \end{array} \right).
\end{eqnarray}
By choosing constant matrix at the unit element of $H_{4}$ as
\begin{eqnarray}\label{45}
E^{+}(e)=\left(
              \begin{array}{cccc}
                0 & 0 & 0 & -1 \\
                0 & 0 & 1 & 0 \\
                0 & 1 & 0 & 0 \\
               -1 & 0 & 0 & 0 \\
              \end{array} \right),
\end{eqnarray}
the original sigma model \eqref{32} is found to be given by
\begin{eqnarray}\label{46}
S=\frac{1}{2}\int d\xi^{+}\wedge d\xi^{-}\Big\{-\partial_+x \partial_-v -\partial_+v \partial_-x
+e^x\Big(\partial_+y \partial_-u+\partial_+u \partial_-y\nonumber\\
+y\partial_+u \partial_-x- y\partial_+x \partial_-u\Big) \Big\},
\end{eqnarray}
such that by rescaling $\kappa=\frac{2\pi}{K}$ in \eqref{22}, the above action will be equivalent to $H_{4}$WZW model. Comparing the above action with the sigma model action of the form \eqref{1}, one can read off the symmetric metric $g_{\mu\nu}$ and antisymmetric tensor field $b_{\mu\nu}$, as follows
\begin{eqnarray}\label{47}
g_{\mu \nu}~=~\left( \begin{tabular}{cccc}
              $0$ & $0$ & $0$ & $-1$\\
              $0$ & $0$ & $e^x$ & $0$ \\
              $0$ & $e^x$ & $0$ & $0$ \\
              $-1$ & $0$ & $0$ & $0$\\
                \end{tabular} \right)~~~~~,~~~~~b_{\mu \nu}~=~\left( \begin{tabular}{cccc}
                              $0$ & $0$ & $-y e^x$ & $0$\\
                              $0$ & $0$ & $0$ & $0$ \\
                              $y e^x$ & $0$ & $0$ & $0$ \\
                              $0$ & $0$ & $0$ & $0$\\
                                \end{tabular} \right).
\end{eqnarray}
\subsubsection{\normalsize The dual model}
To construct the dual sigma model, we parametrize the corresponding Lie group ${A}^{-1,1}_{4,5}.i$ with coordinates $\{{\tilde x},{\tilde y},{\tilde u},{\tilde v}\}$ in the same parametrization \eqref{19}, as
\begin{eqnarray}\label{48}
{\tilde g}\;=\; e^{{\tilde v} {\tilde X}^4} ~ e^{{\tilde u} {\tilde X}^3} ~ e^{{\tilde x} {\tilde X}^1}~ e^{{\tilde y} {\tilde X}^2}.
\end{eqnarray}
Then, from ${\tilde L}_{\pm}=({\tilde g}^{-1}{\partial}_{\pm}~{\tilde g})_{i}~{\tilde X}^{i}$ one can obtain the components of left invariant one-forms ${\tilde L}_{\pm i}$'s as follows
\begin{eqnarray}\label{49}
{\tilde L}_{\pm 1}={\partial}_{\pm}{\tilde x}+\frac{\tilde x}{2}~{\partial}_{\pm}{\tilde v},~~~~~
{\tilde L}_{\pm 2}={\partial}_{\pm}{\tilde y}-\frac{\tilde y}{2}~{\partial}_{\pm}{\tilde v},~~~~~
{\tilde L}_{\pm 3}={\partial}_{\pm}{\tilde u}+\frac{\tilde u}{2}~{\partial}_{\pm}{\tilde v},~~~~~
{\tilde L}_{\pm 4}={\partial}_{\pm}{\tilde v}.
\end{eqnarray}
Moreover, by a direct application of formula \eqref{36}, the matrix ${\tilde a}(\tilde g)$ is obtained as
\begin{eqnarray}\label{50}
{\tilde a}(\tilde g)^{i}\hspace{0cm}_{j}=\left(
              \begin{array}{cccc}
                e^{\frac{-\tilde v}{2}} & 0 & 0 & 0 \\
                0 & e^{\frac{\tilde v}{2}} & 0 & 0 \\
                0 & 0 & e^{\frac{-\tilde v}{2}} & 0 \\
               \frac{\tilde x}{2} & \frac{-\tilde y}{2} & \frac{\tilde u}{2} & 1 \\
              \end{array} \right).
\end{eqnarray}
Then, from \eqref{50} and the following ${\tilde r}$-matrix
\begin{eqnarray}\label{51}
{\tilde r}_{ij}=\left(
              \begin{array}{cccc}
                0 & 0 & 0 & -2 \\
                0 & 0 & -2 & 0 \\
                0 & 2 & 0 & 0 \\
                2 & 0 & 0 & 0 \\
              \end{array} \right),
\end{eqnarray}
and using ${\tilde\sigma}=\frac{\tilde v}{2}$; one can obtain ${\tilde \Lambda}({\tilde g})$ in \eqref{38} as follows
\begin{eqnarray}\label{52}
{\tilde \Lambda}(\tilde g)_{ij}=\left(
              \begin{array}{cccc}
                e^{\frac{-\tilde v}{2}} & 0 & 0 & 0 \\
                0 & e^{\frac{\tilde v}{2}} & 0 & 0 \\
                0 & 0 & e^{\frac{-\tilde v}{2}} & 0 \\
               \frac{\tilde x}{2} & \frac{-\tilde y}{2} & \frac{\tilde u}{2} & 1 \\
              \end{array} \right).
\end{eqnarray}
Finally, by substituting \eqref{50} and \eqref{52} in \eqref{35} and considering ${\tilde E}^{+}(\tilde e)=(E^{+}(e))^{-1}$, the dual sigma model action \eqref{33} takes the following form
\begin{eqnarray*}
{\tilde S}=\frac{1}{2} \int d\xi^+\wedge~d\xi^-\Big\{-e^{\frac{\tilde v}{2}}(\partial_+{\tilde x } \partial_-{\tilde v}+\partial_+{\tilde v } \partial_-{\tilde x})+\frac{1}{3-2e^{\frac{-\tilde v}{2}}}(\partial_+{\tilde y } \partial_-{\tilde u}+{\tilde u}\partial_+{\tilde y } \partial_-{\tilde v}+{\tilde y}\partial_+{\tilde v } \partial_-{\tilde u})
\end{eqnarray*}
\vspace{-6mm}
\begin{eqnarray}\label{53}
\hspace{1cm}~~~~~~+\frac{1}{1-2e^{\frac{-\tilde v}{2}}}(-\partial_+{\tilde u } \partial_-{\tilde y}+{\tilde y}\partial_+{\tilde u } \partial_-{\tilde v}+{\tilde u}\partial_+{\tilde v } \partial_-{\tilde y})-\frac{2{\tilde y}{\tilde u}}{(1-2e^{\frac{-\tilde v}{2}})(3-2e^{\frac{-\tilde v}{2}})}\partial_+{\tilde v } \partial_-{\tilde v}\Big\}.
\end{eqnarray}
By identifying the above action with the dual sigma model of the form \eqref{7'} with ${\tilde\sigma}=\frac{\tilde v}{2}$, symmetric and antisymmetric backgrounds ${\tilde g}_{\mu\nu}$ and ${\tilde b}_{\mu\nu}$ are obtained to be of the form
\begin{eqnarray*}
{\tilde g}_{\mu \nu}~=\zeta\left( \begin{tabular}{cccc}
               $0$ & $0$ & $0$ & $(2-e^{\frac{\tilde v}{2}})(3-2e^{\frac{-\tilde v}{2}})$\\
               $0$ & $0$ & $-1$ & $2{\tilde u}(1-2e^{\frac{-\tilde v}{2}})$\\
               $0$ & $-1$ & $0$ & $2{\tilde y}(1-2e^{\frac{-\tilde v}{2}})$\\
               $(2-e^{\frac{\tilde v}{2}})(3-2e^{\frac{-\tilde v}{2}})$ & $2{\tilde u}(1-2e^{\frac{-\tilde v}{2}})$ & $2{\tilde y}(1-2e^{\frac{-\tilde v}{2}})$ & $-2{\tilde y}{\tilde u}$\\
               \end{tabular}\right),
\end{eqnarray*}
\begin{eqnarray}\label{54}
 {\tilde b}_{\mu \nu}~=\zeta\left( \begin{tabular}{cccc}
               $0$ & $0$ & $0$ & $0$\\
               $0$ & $0$ & $2(1-e^{\frac{-\tilde v}{2}})$ & $-{\tilde u}$\\
               $0$ & $-2(1-e^{\frac{-\tilde v}{2}})$ & $0$ & ${\tilde y}$\\
               $0$ & ${\tilde u}$ & $-{\tilde y}$ & $0$\\
               \end{tabular} \right),
\end{eqnarray}
where $\zeta=\frac{e^{\frac{-\tilde v}{2}}}{(1-2e^{\frac{-\tilde v}{2}})(3-2e^{\frac{-\tilde v}{2}})}$ . Therefore, the complicated action \eqref{53} by the Jacobi-Lie T-duality can be transformed to very simpler action \eqref{46} and vice versa.
\section{\large Conformal invariance of the Jacobi-Lie T-dual sigma models up to the one-loop $\beta$-functions}
The string theory as a conformal invariant two-dimensional sigma model should have conditions on its background matrix. These conformal invariance conditions of a two-dimensional sigma model are nothing but the vanishing of $\beta$-functions, which can be interpreted as equations of motions of the string effective action\cite{callan}. The $\beta$-function equations for the following non-linear sigma model on the target manifold $M$ with symmetric background $G_{\mu\nu}$, antisymmetric tensor field $B_{\mu\nu}$ and dilaton field $\Phi$, as
\begin{equation}\label{55}
S=\frac{1}{2}\int{d\xi^{+} \wedge d\xi^{-}~}{\large \{}({G}_{\mu\nu}+{B}_{\mu\nu})~ \partial_{+}x^{\mu}\partial_{-}x^{\nu}-\frac{R}{4}\Phi(x^{\mu}){\large \}},
\end{equation}
in the lowest order (one-loop and $\alpha'$-correction) are given by \cite{callan}

\begin{eqnarray}\label{56}
{\beta}^{^{G}}_{\mu\nu} =R_{{\mu \nu}}+2{\nabla}_\mu
{\nabla}_\nu \Phi -\frac{1}{4}(H^2)_{\mu \nu},\nonumber\\
{\beta}^{^{B}}_{\mu\nu} ={\nabla}_\lambda\;H_{{\mu \nu}}^{~~\;\lambda} -2({\nabla}_\lambda\Phi)~ H_{{\mu \nu}}^{~~\;\lambda},\nonumber\\
{\beta}^{^{\Phi}} =4({\nabla}\Phi)^2-4{\nabla}^2 \Phi  - {\cal R}+\frac{1}{12} H^2,
\end{eqnarray}
where $R_{\mu\nu}$ and ${\cal R}$ are the Ricci tensor and Ricci scalar of the metric $G_{\mu\nu}$, respectively. In addition, $H_{\mu\nu\lambda}$ is the torsion of the antisymmetric field $B_{\mu\nu}$ with the following relation
\begin{eqnarray}\label{57}
H_{{\mu \nu \lambda}} = \partial_{\mu} B_{\nu \lambda}+\partial_{\nu} B_{\lambda \mu}+\partial_{\lambda} B_{\mu \nu},
\end{eqnarray}
so that, $(H^2)_{\mu \nu} = H_{{\mu \lambda \rho }} H^{{\lambda \rho}}_{~~\nu}$ and $H^2 = H_{{\mu \nu \lambda}} H^{{\mu \nu \lambda}} $.
Indeed, the above $\beta$-functions are the equations of motion for the following effective string action on $m$-dimensional ($m=dim M$) target manifold
\begin{equation}\label{58}
S=\frac{1}{2}\int{d^{m}x}~\sqrt{G}~e^{-2\Phi}~{\large [}{{\cal R}+4({\nabla}\Phi)^2-\frac{1}{12} H^2{\large ]}},
\end{equation}
where $G=det\;G_{\mu\nu}$ \cite{FeTs}.\\
For Jacobi-Lie T-dual sigma models, the action \eqref{55} with zero dilaton field $(\Phi)$ should be converted to two-dimensional sigma model \eqref{1}. For this reason, we must consider the following sigma model (and its dual) for Jacobi-Lie T-dual case
\begin{equation}\label{59}
S=\frac{1}{2}\int{d\xi^{+} \wedge d\xi^{-}~}{\large \{}e^{\sigma}({g}_{\mu\nu}+{b}_{\mu\nu})~ \partial_{+}x^{\mu}\partial_{-}x^{\nu}-\frac{R}{4}\Phi(x^{\mu}){\large \}}.
\end{equation}
Comparing \eqref{55} and \eqref{59}, we should insert
\begin{equation}\label{60}
G_{\mu\nu}=e^{\sigma}g_{\mu\nu},~~~~~B_{\mu\nu}=e^{\sigma}b_{\mu\nu}
\end{equation}
in $\beta$-function equations \eqref{56} for considering conformal invariance in Jacobi-Lie T-dual sigma models. Now, we will try to examine the conformal invariance of the Jacobi-Lie T-dual sigma models \eqref{46} and \eqref{53} up to one-loop order.\\
We know that the original model \eqref{46} as a WZW model is a conformal sigma model. For this model, we obtain Ricci scalar ${\cal R}$ equal to zero and non-constant dilation field as
\begin{equation}\label{61}
\Phi(x^{\mu})={\cal C}+{\cal D} x,
\end{equation}
where ${\cal C}$ and ${\cal D}$ are arbitrary constants. These results were expected, because of the original model as a WZW model is conformal invariance. Now, we will prove the conformal invariance of the dual model \eqref{53}. For this model, we have ${\tilde \sigma}=\frac{\tilde v}{2}$ and
\begin{eqnarray*}
{\tilde G}_{\mu \nu}~=\eta\left( \begin{tabular}{cccc}
               $0$ & $0$ & $0$ & $(2-e^{\frac{\tilde v}{2}})(3-2e^{\frac{-\tilde v}{2}})$\\
               $0$ & $0$ & $-1$ & $2{\tilde u}(1-2e^{\frac{-\tilde v}{2}})$\\
               $0$ & $-1$ & $0$ & $2{\tilde y}(1-2e^{\frac{-\tilde v}{2}})$\\
               $(2-e^{\frac{\tilde v}{2}})(3-2e^{\frac{-\tilde v}{2}})$ & $2{\tilde u}(1-2e^{\frac{-\tilde v}{2}})$ & $2{\tilde y}(1-2e^{\frac{-\tilde v}{2}})$ & $-2{\tilde y}{\tilde u}$\\
               \end{tabular}\right),
\end{eqnarray*}
\begin{eqnarray}\label{62}
 {\tilde B}_{\mu \nu}~=\eta\left( \begin{tabular}{cccc}
               $0$ & $0$ & $0$ & $0$\\
               $0$ & $0$ & $2(1-e^{\frac{-\tilde v}{2}})$ & $-{\tilde u}$\\
               $0$ & $-2(1-e^{\frac{-\tilde v}{2}})$ & $0$ & ${\tilde y}$\\
               $0$ & ${\tilde u}$ & $-{\tilde y}$ & $0$\\
               \end{tabular} \right),
\end{eqnarray}
where $\eta=\frac{1}{(1-2e^{\frac{-\tilde v}{2}})(3-2e^{\frac{-\tilde v}{2}})}$. We obtain that the only non-zero components for the dual model as follows
\begin{equation}\label{63}
{\tilde R}_{\tilde v \tilde v}=-\frac{6e^{-\tilde v}}{(1-2e^{\frac{-\tilde v}{2}})^{2}(3-2e^{\frac{-\tilde v}{2}})^{2}},~~~~~~{\tilde H}_{\tilde y\tilde u\tilde v}=\frac{6+e^{-\tilde v}(16-21 e^{\frac{\tilde v}{2}}-4 e^{-\frac{\tilde v}{2}})}{(1-2e^{\frac{-\tilde v}{2}})^{2}(3-2e^{\frac{-\tilde v}{2}})^{2}},
\end{equation}
such that for this model we have the zero scalar curvature ${\cal \tilde R}=0$ (i.e. the dual target space is also flat). By considering ${\tilde H}_{\tilde y\tilde u\tilde v}$ from \eqref{63}, we can find that $({\tilde H})^{{2}}=0$ and the only non-zero component
$({\tilde H}^{2})_{\tilde v \tilde v}=-2{\Big[}\frac{(e^{-\tilde v}-2)\{6+e^{-\tilde v}(16-21 e^{\frac{\tilde v}{2}}-4 e^{-\frac{\tilde v}{2}})\}}{(1-2e^{\frac{-\tilde v}{2}})^{2}(3-2e^{\frac{-\tilde v}{2}})^{2}}{\Big ]}^{2}$.
Considering the above results, it is straightforward to obtain the dilaton field for the dual model, as
\begin{equation}\label{64}
{\tilde \Phi}({\tilde x}^{\mu})={\cal \tilde C}+2({\cal \tilde D}e^{\frac{\tilde v}{2}}+e^{\frac{-\tilde v}{2}})+\frac{1}{2}\ln{\Big[}\frac{e^{4\tilde v}}{(1-2e^{\frac{-\tilde v}{2}})(3-2e^{\frac{-\tilde v}{2}})}{\Big ]}-\frac{1}{6}e^{-\tilde v}.
\end{equation}
where ${\cal \tilde C}$ and ${\cal \tilde D}$ are arbitrary constants. Note that the singular point in the action \eqref{53} (by vanishing of the Ricci scalar ${\cal \tilde R}$, $\tilde R_{\mu\nu}\tilde R^{\mu\nu}$ and Kretschman scalar $\tilde R_{\mu\nu\rho\gamma}\tilde R^{\mu\nu\rho\gamma}$) is not an essential point and it can be removed by a coordinate transformation.
\section{\large Discussion and Conclusion}
We showed that the WZW model on the Heisenberg Lie group $H_{4}$ has the Jacobi-Lie symmetry with four dual Lie groups where dual Lie group $A_{2}\oplus 2A_{1}$ (found by Poisson-Lie symmetry) is one of them. Then, we constructed Jacobi-Lie T-dual sigma models on Jacobi-Lie bialgebra $((h_{4},0),({\cal A}^{-1,1}_{4,5}.i,\frac{1}{2}X_{4}))$ with functions $\sigma=0$ and ${\tilde{\sigma}}=\frac{\tilde v}{2}$, and showed that the original sigma model is equivalent to the $H_{4}$WZW model. Also, we studied the conformality of the dual model related to $((h_{4},0),({\cal A}^{-1,1}_{4,5}.i,\frac{1}{2}X_{4}))$ at the one-loop order and found its dilaton fields.

\bigskip

\noindent {\bf Acknowledgments:}
We would like to express our gratitude to S. Hosseinzadeh and R. Gholizadeh-Roshanagh for their useful comments.


\end{document}